\documentclass[11pt]{article}
\usepackage{a4wide}
\usepackage{amssymb}
\usepackage{dsfont}
\begin{document}
{\renewcommand{\thefootnote}{\fnsymbol{footnote}}
\hfill  IGC--10/9--2\\
\mbox{}\hfill ITP-UU-10/35\\
\mbox{}\hfill SPIN-10/30\\
\vspace{1.5cm}
\begin{center}
{\LARGE  An effective approach to the problem of time}\\
\vspace{2em}
Martin Bojowald,\footnote{e-mail address: {\tt bojowald@gravity.psu.edu}}$^1$
Philipp A H\"ohn\footnote{e-mail address: {\tt p.a.hohn@uu.nl}}$^{2,1}$, and
Artur Tsobanjan\footnote{e-mail address: {\tt axt236@psu.edu}}$^1$
\\
\vspace{1em}
$^1$Institute for Gravitation and the Cosmos,\\
The Pennsylvania State
University,\\
104 Davey Lab, University Park, PA 16802, USA\\
\vspace{1em}
$^2$ Institute for Theoretical Physics,\\
 Universiteit Utrecht,\\
Leuvenlaan 4, NL-3584 CE Utrecht, The Netherlands

\vspace{1.5em}
\end{center}
}

\setcounter{footnote}{0}

\begin{abstract}
 A practical way to deal with the problem of time in quantum cosmology
 and quantum gravity is proposed. The main tool is effective
 equations, which mainly restrict explicit considerations to
 semiclassical regimes but have the crucial advantage of allowing the
 consistent use of local internal times in non-deparameterizable
 systems. Different local internal times are related merely by gauge
 transformations, thereby enabling relational evolution through turning points of non-global internal times. The main consequence of the local nature
 of internal time is the necessity of its complex-valuedness,
 reminiscent of but more general than non-unitarity of evolution
 defined for finite ranges of time. By several general arguments, the
 consistency of this setting is demonstrated. Finally, we attempt an outlook on the nature of time in highly quantum regimes.
 The focus of this note is on conceptual issues.
\end{abstract}

\section{Introduction}

The problem of time \cite{Kuc1,Ish,Rovbook,anderson} arises in quantum
gravity because the dynamics of a generally covariant theory is fully
constrained, without a true Hamiltonian generating evolution with
respect to a distinguished time. Moreover, the relational interpretation of
evolution is complicated by the {\it global time problem}, the fact that a globally valid choice of
internal time is difficult to find and may not exist. For specific
matter systems, such as a free massless scalar field or pressureless
dust, deparameterizations with a matter clock can be performed, but
these models seem rather special. In order to evaluate the dynamics of
quantum gravity and derive potentially observable information from
first principles, the problem of time must be overcome at a general
level without requiring specific adaptations.

For most applications of quantum gravity related to potential
observables, semiclassical evolution is sufficient or at least provides
a great deal of information. One may then hope that such a situation
makes tackling the problem of time more feasible since this
problem does not play a handicapping role classically; or at least a
dedicated analysis of semiclassical evolution should provide insights
which may help in attacking the problem in full generality. In this
article, we use the effective approach to quantum constraints for finite dimensional systems
developed in \cite{EffCons,EffConsRel} in the context of the problem
of time, and we propose a practical solution employing local rather
than global internal times.

The concept of using an internal time even if it is not well-defined
globally and switching to a new time when this becomes necessary, is
easy to apply classically and very reminiscent of the concept of local
coordinates on a manifold. Obviously, if an atlas of local internal
times can be made to work even in quantum gravity, this concept has
the potential of leading to a solution to the problem of time. From
the point of view of state evolution, local internal times in quantum
systems have been discussed, for instance, in \cite{Rovmod}, but no
clear solution as regards evolution has been reached. (Specific
constructions in the Bianchi I model have been suggested in
\cite{PhysEvolBI}.)  The main problem is not unexpected: If time is
defined only for a finite range, unitary evolution of states cannot be
realized. While classical evolution with respect to a local internal
time is unproblematic all the way to and --- by patching --- even
through its turning point, non-unitary quantum evolution is in danger
of producing meaningless results long before the end of one local
internal time is reached. Moreover, it is not clear how to define
quantum observables in such a situation. The technical Hilbert-space
issues related to these conceptual problems in the context of time and
evolution seem too difficult to be resolved even in simple models, let
alone in a practical manner for generic situations in quantum gravity.

At this stage, effective techniques which describe a quantum system
and its dynamics via expectation values and moments assigned by a state
become important. While these tools describe the full quantum system
--- usually approximately and for specific classes of states --- for
many purposes they produce equations that can be treated by
well-known classical procedures. As we will see, the new viewpoint
also sheds light on issues of time and especially the use of local
internal times in quantum systems. Instead of non-unitarity of the
evolution, we will encounter the need to use complex-valued times;
but in contrast to problems with evolution of states in a Hilbert
space, the effective evolution of expectation values and moments
with respect to complex time can easily be made sense of. These
features of complex time in the effective formulation, as well as
analogous ones that lie more hidden in standard treatments of state
evolution, are the focus of this note and, in particular, of
Section~\ref{s:Complex}.  Moreover, switching local internal times
within the effective treatment can be handled consistently and
requires nothing more than a gauge transformation. Toward the end of
this article, we will dare an outlook on the nature of time in
non-semiclassical, highly quantum states.

\section{Effective constraints}\label{seceffcon}

We consider a quantum system subject to a single constraint operator
$\hat{C}$ playing the role of a Hamiltonian constraint. Physical
states thus satisfy $\hat{C}|\psi\rangle=0$. Assumptions about the
spectrum of $\hat{C}$ will not be made; in particular, effective
techniques work for zero in the discrete as well as the continuous
part of the spectrum of constraint operators.

For a systematic derivation of effective descriptions for canonical
quantum theories we parameterize states by expectation values and
moments rather than wave functions or density matrices: For several
pairs of canonical degrees of freedom $(q_1,p_1;q_2, p_2;\ldots
;q_n,p_n)$, we use the expectation values $\langle\hat{q_i}\rangle$
and $\langle\hat{p_i}\rangle$, $i=1, \ldots n$, together with the
moments
\[
 \Delta(q_1^{a_1}p_1^{b_1} q_2^{a_2}p_2^{b_2} \ldots):=\langle(\hat{q}_1-\langle\hat{q}_1\rangle)^{a_1}
(\hat{p_1}-\langle\hat{p_1}\rangle)^{b_1} \ldots \rangle_{\rm Weyl}
\]
(ordered totally symmetrically and defined for $\sum_i (a_i+b_i)\geq
2$) as a complete description of states. (For instance,
$\Delta(q_i^2)=(\Delta q_i)^2$ is the position fluctuation of the
$i$-th coordinate.)

The manifold spanned by expectation values and moments carries a
phase-space structure defined by the Poisson bracket
\[
 \{\langle\hat{A}\rangle,\langle\hat{B}\rangle\}=
\frac{\langle[\hat{A},\hat{B}]\rangle}{i\hbar}
\]
for any pair of operators $\hat{A}$ and $\hat{B}$, extended to the
moments using the Leibniz rule and linearity.  If there is a true
Hamiltonian, it follows from the Heisenberg equation that the
Hamiltonian flow of expectation values and moments is generated by the
quantum Hamiltonian
$H_Q(\langle\hat{q}\rangle,\langle\hat{p}\rangle,\Delta(\cdots))=
\langle\hat{H}\rangle$.

For a constraint, the expectation values and moments assigned by physical
states must satisfy $\langle\hat{C}\rangle=0$ as a constraint function on the
quantum phase space, but also
\[
 C_{\rm pol}:= \langle (\widehat{\rm pol}-\langle\widehat{\rm
   pol}\rangle) \hat{C}\rangle=0
\]
for all polynomials $\widehat{\rm pol}$ in basic operators must
vanish.  This set forms infinitely many first-class constraints for
infinitely many variables.  Notice the ordering: while moments are
defined via a totally symmetric ordering of operators, this is not
done for the quantum constraints; otherwise they would not be first
class \cite{EffCons}. As a consequence, some of the quantum
constraints take complex values. As already shown for
deparameterizable systems \cite{EffCons,EffConsRel}, this
complex-valuedness is not problematic. It simply reflects the fact
that quantum constraints are formulated on the states that take
values on the full algebra of kinematical operators, not all of
which correspond to physical observables once the constraint is
implemented. Hence, the kinematical moments that appear in the
expressions of constraints need not be restricted to real values.
After implementing the constraints, reality conditions can be
imposed on the physical expectation values and moments --- the Dirac
observables of the constrained system --- and contact with the
physical Hilbert space can be made.

The set of infinitely many constraints for infinitely many variables
is directly tractable by exact means if the constraints decouple
into finite sets, a situation realized only for constraints linear
in canonical variables. For more interesting cases one must use
approximations that allow one to reduce the system by ignoring
subdominant terms. The prime example for such an approximation is
the semiclassical expansion, corresponding to states whose moments
of high orders are suppressed compared to expectation values and
lower-order moments. The simple and still rather general assumption
$\Delta(q^ap^b)=O(\hbar^{(a+b)/2})$ allows one to arrange all
contributions to the constraints in such a way that to any given
finite order in $\hbar$ only finitely many constraints contribute,
allowing one to solve for all physical moments up to the order
considered. This semiclassicality assumption will be used in the
following discussions (except in parts of section \ref{hq}). We note that this restriction on the states
considered still leaves a large class, much larger, certainly, than
the common specification of a Gaussian state (whose moments are
completely fixed by specifying just the second-order moments) would
allow.

\subsection{Example: ``Relativistic'' harmonic oscillator}\label{secrelho}

Consider $\hat{C}=\hat{p}_t^2 - \hat{p}_{\alpha}^2 -
\hat{\alpha}^2$. To second order in the moments, we obtain the quantum
constraints \cite{EffConsComp}
\begin{eqnarray}
C&=& \langle\hat{p}_t\rangle^2 - \langle\hat{p}_{\alpha}\rangle^2 -
\langle\hat{\alpha}\rangle^2 + (\Delta p_t)^2 - (\Delta
p_{\alpha})^2 - (\Delta \alpha)^2 \label{eq:c_funct1} \\ C_{t}& =&
2\langle\hat{p}_t\rangle \Delta(t p_t) + i\hbar
\langle\hat{p}_t\rangle - 2\langle\hat{p}_{\alpha}\rangle \Delta(t
p_{\alpha}) - 2\langle\hat{\alpha}\rangle
\Delta(t\alpha) \\
C_{p_t}& =&
2\langle\hat{p}_t\rangle (\Delta p_t)^2 -
2\langle\hat{p}_{\alpha}\rangle  \Delta(p_tp_{\alpha})
-2\langle\hat{\alpha}\rangle\Delta(p_t\alpha)\\
C_{\alpha} & =& 2\langle\hat{p}_t\rangle \Delta(p_t\alpha) -
2\langle\hat{p}_{\alpha}\rangle \Delta(\alpha p_{\alpha})
- i\hbar \langle\hat{p}_{\alpha}\rangle-
2\langle\hat{\alpha}\rangle(\Delta \alpha)^2 \\
C_{p_{\alpha}}& =& 2\langle\hat{p}_t\rangle \Delta(p_tp_{\alpha}) -
2\langle\hat{p}_{\alpha}\rangle(\Delta p_{\alpha})^2
-2\langle\hat{\alpha}\rangle \Delta(\alpha p_{\alpha}) + i\hbar
\langle\hat{\alpha}\rangle\, . \label{eq:c_funct2}
\end{eqnarray}
These constraint functions are first-class to order $\hbar$\ and
therefore generate gauge transformations. This is a key difference
between the standard Dirac constraint quantization at the
Hilbert-space level and the effective approach: after solving the
quantum constraint in the former method all gauge flows are absent
in the physical Hilbert space, whereas solving the constraints at the
effective level does not immediately lead to gauge invariance. One
way of understanding this difference is to note that the states of
the physical Hilbert space assign expectation values only to the
physical Dirac observables, while in the effective approach, states
assign expectation values to all kinematical variables, which in
general are subject to gauge even classically. Gauge invariance at
the effective level is only achieved by constructing effective Dirac
observables, at which point the number of (true) degrees of freedom
in the two approaches coincides.

Following
\cite{EffCons,EffConsRel} we fix the gauge that for
deparameterizable systems corresponds to the evolution of
$\hat{\alpha}$\ and $\hat{p}_{\alpha}$\ in $\hat{t}$, by setting
fluctuations of the latter to zero
\begin{eqnarray}\label{gauge}
(\Delta t)^2 = \Delta(t \alpha) = \Delta (t p_{\alpha}) = 0\, .
\end{eqnarray}
Imaginary contributions to the constraints arise, which require some
of the moments to take complex values. For instance, $\Delta(tp_t)=
-\frac{1}{2}i\hbar$ if one imposes the above gauge choice. All the
gauge-fixed moments refer to $t$ which, when chosen as time in this
deparameterizable system, is not represented as an operator and does
not generate physical moments. The gauge-dependence or
complex-valuedness of these moments is, therefore, not a problem. In
fact, the complex-valuedness of the moments guarantees that
generalized uncertainty relations are respected even if some
fluctuations vanish. In particular, the gauge (\ref{gauge}) actually
leads to a saturation of the (generalized) uncertainty relation
$(\Delta t)^2(\Delta p_t)^2-(\Delta(tp_t))^2\geq\hbar^2/4$.

Moments not involving time or its momentum, on the other hand,
should have a physical analog taking strictly real values. That this
is the case can be inferred from the fact that the second-order
effective quantum system can be deparameterized by solving for
$\langle\hat{p}_t\rangle=\pm H_Q$ with the quantum Hamiltonian
\[
H_Q = \sqrt{\langle\hat{p}_{\alpha}\rangle^2 + \langle\hat{\alpha}\rangle^2}
\Biggl( 1 + \frac{ \langle\hat{\alpha}\rangle^2(\Delta p_{\alpha} )^2 -
2\langle\hat{\alpha}\rangle \langle\hat{p}_{\alpha}\rangle
\Delta(\alpha p_{\alpha}) +
\langle\hat{p}_{\alpha}\rangle^2
(\Delta \alpha)^2}{2(\langle\hat{p}_{\alpha}\rangle^2
+ \langle\hat{\alpha}\rangle^2)^2} \Biggr)\, .
\]
Solving the Hamiltonian equations of motion for
$\langle\hat{\alpha}\rangle(t)$,
$\langle\hat{p}_{\alpha}\rangle(t)$, $\Delta(\cdots)(t)$ gives the
Dirac observables of the constrained system, on which reality can
easily be imposed simply by requiring real initial values at some
$t$. Although there is a true operator $\hat{t}$ at the kinematical
level, its expectation value does not appear in the effective
constraints. In the final equations of motion for the physical
evolving observables, it just appears as an evolution parameter.

\subsection{Non-deparameterizable systems}

\begin{quote}
{\em Zeit ist das, was man an der Uhr abliest. (Time is what you read off the clock.)}

\mbox{}\hfill {\sc Albert Einstein}
\end{quote}

We now turn to systems in which no global internal time exists,
realized, for instance, in the presence of a time-dependent
potential in the constraint operator. To be specific, by a
non-global internal time we mean a clock variable whose equal-time
surfaces may be intersected more than once or not at all by a
classical trajectory; the clock will, therefore, encounter one or
more extrema in the course of classical evolution. Such systems do
occur in the context of general relativity, one simple example being
a $k=1$ FRW universe filled with a massive scalar field. While
classically there is no profound problem with non-global clocks
since, in principle, one can always revert to a time coordinate,
i.e., the gauge parameter along the flow of the Hamiltonian
constraint, the time coordinate is absent in the quantum theory,
where physical states are automatically gauge invariant. In these
situations evolution with respect to local internal times is
required. A coherent state of the corresponding quantum system which
is peaked on a classical trajectory, must then decay beyond the
classical turning point of the local clock, so that the quantum
evolution with respect to such clocks appears non-unitary.

Concerning evolution, the choice and corresponding notion of time is
associated to the clock which we are employing. As indicated already
by the deparameterizable system, the choice of clock within the
effective treatment is implemented by gauge fixing: moments involving
time (such as $\Delta t$) vanish (or take other prescribed and
possibly complex values). The choice of time and clock is, thus,
closely related to, and, in fact, nothing more than a gauge choice; we
will refer to the choice as a {\it Zeitgeist}.

Applying such gauge conditions to non-global clocks, we encounter a
striking feature: a consistent solution of the constraints and
equations of motion requires complex-valued time.  Interestingly,
this novel feature can be implemented in a self-consistent manner
and leads us to well-defined effective dynamics in a local time away
from its classical extremal points. However, as we evolve closer
towards these points with respect to the local clock, spreads and
other moments diverge and become singular in the gauge corresponding
to our choice of time \cite{BHTlongarticle}; this gauge and, thus, the choice of time
becomes incompatible with the semiclassical expansion of moments. The apparent non-unitarity in a non-global
time at the state level translates into the eventual breakdown of
the corresponding gauge in the effective formalism. Fortunately,
this ailment can be cured if we can find a new clock which is
locally better-behaved where our original time variable becomes
inadequate. Since the choice of time is nothing more than a gauge
choice, we can switch clocks by a suitable gauge transformation,
which in the effective treatment is generated by the constraint
functions (such as~(\ref{eq:c_funct1})-(\ref{eq:c_funct2})), in
close analogy with classical constrained systems. Such
transformations were, indeed, found explicitly in toy models and,
hence, those systems could be evolved along semiclassical
trajectories through the extremal points of local times, by
``temporarily'' switching to different clock functions
\cite{BHTlongarticle}. In this way we attempt to reconstruct
effectively a coherent physical state.

As regards the relational interpretation, we emphasize that each
choice of a clock and the corresponding gauge comes with a different
description of the system --- its own Zeitgeist.  Specifically, the
moments of the kinematical operator used to measure time are fixed
by the gauge conditions --- only its expectation value remains free;
the conjugate momentum of time is entirely eliminated through the
constraint functions. Therefore, in this description, neither
operator could correspond to a physical variable, which could be
meaningfully turned into a physical (relational) observable.
Changing the clock and, thus, the notion of time, brings about a
significant shift in perspective regarding the physical variables:
the old clock and its conjugate momentum become physical in the new
regime, while the newly chosen clock is relegated to the status of a
parameter and its conjugate variable is altogether eliminated
through constraints. Moreover, the accompanying gauge changes yield
jumps of order $\hbar$ in physical correlations
\cite{BHTlongarticle}. This has an important implication for
(quantum) relational observables for non-deparameterizable systems,
namely, one cannot construct relational observables which are valid
for values of the relational clock near its turning points. In those
regions we are forced to use a different clock and, therefore, to
evolve a truly \emph{different} set of ``effective local relational
observables''. Trajectories for local relational observables in a
new time, consequently, do not directly continue the preceding ones
in the old time and leave a gap, although, nonetheless, consistently
transporting along relational initial data. Time is of a local
nature here and so is the relational concept of evolution.

Since we are dealing with a first-class constrained system, the
concept of observables as gauge-invariant functions on phase space is
still valid. What becomes limited in the absence of global internal
times is the usual notion of evolving observables
\cite{Rovbook,Rovmod,GeomObs,Bianca1,Haj1}. In particular, if local
clocks have maximal or minimal values along classical trajectories,
these extremal values typically vary from orbit to orbit. It may be
the case for classical trajectories in some systems,\footnote{For
instance, systems with closed orbits.} that sets of values (or even
every value) of a given local clock lie beyond the maximal (or
minimal) clock value allowed by the given classical orbit. Relational
observables evolving in such a non-global clock are generally
multi-valued and become complex beyond the extremal points, indicating
that the system with given initial data will never reach such phase
space points. Hence, the quantum version of a relational (Dirac)
observable referring to this clock can, in principle, be a
well-defined operator\footnote{Or multiple operators if the relational
observable is multi-valued.} on the physical Hilbert-space, but will,
in general, yield complex expectation values in a physical inner
product, thus, failing to be a self-adjoint operator on
$\mathcal{H}_{\rm phys}$ (see also \cite{Rovmod} on this issue). On the
other hand, in a given Zeitgeist at the effective level one may
formally compute expectation values of evolving observables, but a
Zeitgeist is only rarely permanent. When it changes near a turning
point of the local internal time, a different set of local relational
observables is required. To be precise, by local relational
observables at the effective level we mean correlations of expectation
values and moments with the expectation value of the local clock
variable, evaluated in its corresponding Zeitgeist. In this article,
we call such local relational observables of the effective formalism
computed with respect to a Zeitgeist ``fashionables''; they constitute
the complete physical information of interest about the system as long
as the Zeitgeist remains intact, but may fall out of fashion when the
Zeitgeist changes. The notion of a fashionable is, therefore,
state-dependent, in contrast to the operator version of a quantum
relational observable, as in different semiclassical states a given
Zeitgeist is generally valid for different ranges of the associated
local clock. Fashionables become invalid when the associated Zeitgeist
/ choice of time fails on approach to an extremal point of the local
clock and, therefore, before the above mentioned issue of
complex-valued correlations could set in. Fashionables, thus, reflect
the local nature of relational quantum evolution and, by being
state-dependent, are somewhat closer to physical interpretation.

By analogy, we will also refer to expectation values of operators in Hilbert-space representations, obtained via local deparametrizations as discussed in section \ref{schrodreg}, as fashionables. It should be noted that in deparameterizable
systems, where the Zeitgeist of the global clock is defined for its entire
range, fashionables become globally valid and coincide with the expectation values of the standard operator versions of relational Dirac observables, obtained via deparametrization in the Dirac procedure.

\section{Complex time}
\label{s:Complex}

Several independent arguments indicate that internal time should be
considered complex in non-deparameterizable systems. The full strength
of this conclusion can be grasped only in the effective approach, which
we will consider first, but it can be seen to arise also in
Hilbert-space treatments.

\subsection{Effective constraints and complex time}
\label{s:EffComplex}

To be specific, we consider effective constraints for a relativistic
particle in an arbitrary time-dependent potential $V(t,q)$. We will
show that the imaginary contribution to $\langle \hat{t} \rangle$\
is insensitive to the explicit form of $V$. Below, primes at the
potential will refer to its partial $q$-derivatives and dots to
its partial $t$-derivatives. We make use of the effective
constraints
\begin{eqnarray}\label{Cnon}
C&=&\langle\hat{p}_t\rangle^2-\langle\hat{p}\rangle^2+ (\Delta
p_t)^2-(\Delta p)^2+V(\langle\hat{t}\rangle, \langle\hat{q}\rangle)
+ {\textstyle\frac{1}{2}} \ddot{V}(\langle\hat{t}\rangle, \nonumber
\langle\hat{q}\rangle)(\Delta t)^2
\\ && + {\textstyle\frac{1}{2}} V''(\langle\hat{t}\rangle,
\langle\hat{q}\rangle)(\Delta q)^2 + \dot{V}'(\langle\hat{t}\rangle,
\langle\hat{q}\rangle)\Delta(tq) \\ C_t &=&
2\langle\hat{p}_t\rangle\Delta(tp_t)+ i\hbar \langle\hat{p}_t\rangle
-2\langle \hat{p} \rangle \Delta(tp)+\dot{V}(\langle\hat{t}\rangle,
\langle\hat{q}\rangle) (\Delta t)^2 + V'(\langle\hat{t}\rangle,
\langle\hat{q}\rangle) \Delta (tq)\\ C_{p_t}&=&
2\langle\hat{p}_t\rangle(\Delta p_t)^2-
2\langle\hat{p}\rangle\Delta(p_tp)+ \dot{V}(\langle\hat{t}\rangle,
\langle\hat{q}\rangle)(\Delta(tp_t)-{\textstyle\frac{1}{2}}i\hbar)
+ V'(\langle\hat{t}\rangle, \langle\hat{q}\rangle) \Delta(p_t q) \\
C_q &=& 2 \langle \hat{p}_t \rangle \Delta(p_tq)- 2\langle \hat{p}
\rangle \Delta(qp)-i\hbar \langle \hat{p} \rangle +
\dot{V}(\langle\hat{t}\rangle, \langle\hat{q}\rangle) \Delta(tq) +
V'(\langle\hat{t}\rangle, \langle\hat{q}\rangle)(\Delta q)^2 \\
C_p &=& 2 \langle \hat{p}_t \rangle \Delta (p_t p) - 2\langle
\hat{p} \rangle (\Delta p)^2 + \dot{V}(\langle\hat{t}\rangle,
\langle\hat{q}\rangle)\Delta(tp) + V'(\langle\hat{t}\rangle,
\langle\hat{q}\rangle)(\Delta(qp)-{\textstyle\frac{1}{2}}i\hbar)\,.
\end{eqnarray}
To implement $\langle\hat{t}\rangle$ as local internal time, no
$t$-moments should be present. The Zeitgeist $(\Delta
t)^2=\Delta(tq)=\Delta(tp)=0$ should, thus, be a suitable way to fix
the gauge.  We first infer $\Delta(tp_t)=-\frac{1}{2}i\hbar$ from
$C_t=0$. Then $C_{p_t}$ implies
\[
(\Delta p_t)^2 = \frac{\langle \hat{p} \rangle}{\langle \hat{p}_t
\rangle} \Delta (p_tp) + \frac{ i\hbar \dot{V}
(\langle\hat{t}\rangle, \langle\hat{q}\rangle)}{2 \langle \hat{p}_t
\rangle} - \frac{V'(\langle\hat{t}\rangle, \langle\hat{q}\rangle)
}{2\langle \hat{p}_t\rangle}\Delta(p_tq)\,.
\]
Eliminating $\Delta(p_tp)$\ and $\Delta(p_tq)$\  in the expression
above using $C_p$\ and $C_q$, respectively, yields
\[
 (\Delta p_t)^2= \frac{\langle\hat{p}\rangle^2}{\langle\hat{p_t}\rangle^2}
 (\Delta p)^2+
 \frac{i\hbar\dot{V}(\langle\hat{t}\rangle,\langle\hat{q}\rangle)}{2\langle\hat{p}_t\rangle} + \frac{V'(\langle\hat{t}\rangle,
\langle\hat{q}\rangle)^2}{4\langle \hat{p}_t \rangle^2}(\Delta q)^2 -
\frac{\langle \hat{p} \rangle V'(\langle\hat{t}\rangle,
\langle\hat{q}\rangle)}{\langle \hat{p}_t\rangle^2} \Delta (qp)\,
\]
and we finally obtain the alternative expression
\begin{eqnarray} \label{Cimag}
 C&=&\langle\hat{p}_t\rangle^2-\langle\hat{p}\rangle^2+
 \frac{\langle\hat{p}\rangle^2-
 \langle\hat{p}_t\rangle^2}{\langle\hat{p}_t\rangle^2}(\Delta p)^2 +
 \left( \frac{V''(\langle\hat{t}\rangle,
\langle\hat{q}\rangle)}{2} + \frac{V'(\langle\hat{t}\rangle,
\langle\hat{q}\rangle)^2}{4\langle \hat{p}_t \rangle^2} \right)
(\Delta q)^2 \nonumber \\ &&- \frac{\langle \hat{p} \rangle
V'(\langle\hat{t}\rangle, \langle\hat{q}\rangle)}{\langle
\hat{p}_t\rangle^2} \Delta (qp)+
 \frac{i\hbar\dot{V}(\langle\hat{t}\rangle,\langle\hat{q}\rangle)}{2\langle\hat{p}_t\rangle} + V(\langle\hat{t}\rangle, \langle \hat{q} \rangle)
\end{eqnarray}
for the constraint $C=\langle\hat{C}\rangle$ on the space on which
$C_t$, $C_{p_t}$, $C_q$\ and $C_p$ are solved and the Zeitgeist is chosen as
above. One may solve (\ref{Cimag}) for $\langle\hat{p}_t\rangle=H_Q$
as the quantum Hamiltonian for time $t$, and compute fashionables
via the evolution it generates. Here, we are mainly concerned with
properties of $\langle\hat{t}\rangle$.

In (\ref{Cimag}), terms not involving $V$\ and its derivatives
should be real-valued: $\langle\hat{p}\rangle$ and $(\Delta p)^2$
(as well as $\langle \hat{q} \rangle$, $\Delta(qp)$\ and $(\Delta
q)$) are physical variables in this Zeitgeist which ought to be
converted into (real-valued) fashionables by solving the equations
of motion, and $\langle\hat{p}_t\rangle$ can be interpreted
physically as the local energy value which is not conserved with a
time-dependent potential but has a clear meaning. When the
constraint is satisfied, we thus determine the imaginary part of
$\langle\hat{t}\rangle$ from the equation
\begin{equation} \label{Vtime}
{\rm Im} \left(\left( \frac{V''(\langle\hat{t}\rangle,
\langle\hat{q}\rangle)}{2} + \frac{V'(\langle\hat{t}\rangle,
\langle\hat{q}\rangle)^2}{4\langle \hat{p}_t \rangle^2} \right)
(\Delta q)^2 \nonumber - \frac{\langle \hat{p} \rangle
V'(\langle\hat{t}\rangle, \langle\hat{q}\rangle)}{\langle
\hat{p}_t\rangle^2} \Delta (qp)+
\frac{i\hbar\dot{V}(\langle\hat{t}\rangle,\langle\hat{q}\rangle)}{2\langle\hat{p}_t\rangle}
 + V(\langle\hat{t}\rangle, \langle \hat{q} \rangle) \right)=0\,,
\end{equation}
which, in general, can be difficult to solve and does not seem to give
rise to a simple, universal, potential-independent imaginary part of
$\langle\hat{t}\rangle$. For semiclassical states, however, to which
this approximation of effective constraints refers anyway, we Taylor
expand the potential in the imaginary term, expected to be of the order $\hbar$,\[
V(\langle\hat{t}\rangle, \langle \hat{q} \rangle)=V({\rm
Re}\,\langle\hat{t}\rangle+i\,{\rm Im}\,\langle\hat{t}\rangle,
\langle \hat{q} \rangle)= V({\rm Re}\,\langle\hat{t}\rangle, \langle
\hat{q} \rangle)+i\,{\rm Im}\,\langle\hat{t}\rangle\, \dot{V}({\rm
Re}\,\langle\hat{t}\rangle,\langle \hat{q} \rangle)+ O(({\rm
Im}\,\langle\hat{t}\rangle)^2)\,,
\]
while
derivatives of the potential are expanded in an identical manner. To
order $\hbar$, only the last two terms of~(\ref{Vtime}) possess an
imaginary part. The imaginary contribution to $C$, which is
constrained to vanish independently of the real part of $C$, is then
given by $\frac{1}{2}i\hbar\,\dot{V}({\rm Re}\,\langle\hat{t}\rangle,
\langle \hat{q} \rangle )/\langle\hat{p}_t\rangle+
i\,\dot{V}({\rm Re}\,\langle\hat{t}\rangle, \langle \hat{q} \rangle)\,
{\rm Im}\,\langle\hat{t}\rangle+O(\hbar^{3/2})=0$. Thus, it
immediately follows that
\begin{equation}\label{imt}
 {\rm Im}\,\langle\hat{t}\rangle= -\frac{\hbar}{2\langle\hat{p}_t\rangle}\,.
\end{equation}

As the derivation shows, this imaginary contribution to time is a
universal result, independent of the potential. In particular, time
must be complex, although with its imaginary part determined by
phase-space variables, it still provides a 1-dimensional flow.

There are old and well-known arguments in quantum mechanics saying
that time cannot be a self-adjoint operator, for it would be conjugate
to an energy operator bounded from below for stable systems. Since a
self-adjoint time operator would generate unitary shifts of energy by
arbitrary values, a contradiction to the lower bound would be
obtained. The result obtained here looks similar at first sight --- a
non-self-adjoint time operator could, certainly, lead to complex
time-expectation values --- but it is more general. In the present
example, we are using an arbitrary potential which does not
necessarily provide a lower bound for energy. The usual arguments
about time operators, thus, do not apply; instead, our conclusions are
drawn directly from the fact that we are dealing with a time-dependent
potential in a non-deparameterizable system. (For time-independent
potentials, $\langle\hat{t}\rangle$ does not appear in the effective
constraints and can consistently be chosen real. The time dependence
is, therefore, crucial for the present discussion.)

The value of the imaginary part of time is directly related to the
Zeitgeist as we have used it in the derivation. Were we to change
time, which is often required in non-deparameterizable systems which
lack a global internal time, we change the Zeitgeist. Accordingly,
the accompanying gauge transformation must transfer the imaginary
contribution from the old time to the new one since the above argument will also hold for the new clock in its corresponding gauge. This can also be demonstrated explicitly in toy models \cite{BHTlongarticle}.

By implementing changes of time as mere gauge transformations in a
first-class system of constraints, we solve the multiple-choice
aspect of the problem of time. Since the effective approach works at
an algebraic level, rather than directly with Hilbert-space
representations, the Hilbert-space aspect of the problem of time is
bypassed as well. In fact, one may view different gauge choices of
the effective constrained system as different representations that
would normally be realized at the Hilbert-space level. The only
price to pay is that we must deal with complex time, which may be
unfamiliar but does not pose any additional difficulties. In the
remainder of this section we will show that the same value of the
imaginary part can be seen to arise not only in the effective
approach.

\subsection{Schr\"odinger regime for relativistic systems}\label{schrodreg}

In a Dirac-type quantization the main difficulty is usually to
determine a physical inner product with physical evolution, for which
no systematic treatment exists in the case of non-deparameterizable
systems.\footnote{Generalizations of Klein--Gordon type physical inner
products have been suggested based on the notion of asymptotic
positive-frequency solutions \cite{WaldTime}. Another method is based
on spectral decomposition \cite{Master}. In those cases, defining
physical evolution, especially through turning points of local
internal times, remains a challenge.}  Moreover, it seems difficult to
shed light on the origin of imaginary contributions to time from this
perspective since there is normally no clock operator defined on the
physical Hilbert space, whose physical expectation value one could
compute. Instead of the physical inner product associated with the
relativistic system, we will consider a Schr\"odinger equation which
linearizes the relativistic equation in the momentum of internal time.

A relativistic constraint equation of the form
\begin{eqnarray}\label{WdW}
\left(\hat{p}_t^2-\hat{H}^2(\hat{t},\hat{q},\hat{p})\right)\psi(q,t)=0\,,
\end{eqnarray}
where $\hat{H}^2$ is a positive operator at least on a subset of
states, is in general not equivalent to the Schr\"odinger equation
\begin{eqnarray}\label{schrod1}
\left(-i\hbar\partial_t+\hat{H}(t,\hat{q},\hat{p})\right)\psi(q,t)=0\,,
\end{eqnarray}
as it would be in the case of a time-independent Hamiltonian for
positive-frequency solutions.
Solutions to (\ref{schrod1}) rather satisfy the relativistic version
\begin{eqnarray}\label{doubleschrod}
-\hbar^2\partial_t^2\psi=\hat{H}^2\psi+i\hbar\partial_t\hat{H}\psi
\end{eqnarray}
of the constraint.

In this comparison, we implicitly assume, however, that $t$ refers to the same time
variable in both cases, and, in particular, that it always takes real
values. In (\ref{schrod1}), $t$ is a time parameter not associated
with any operator and it would be difficult to justify it taking
complex values. In (\ref{WdW}), however, $t$ is an internal variable
and quantized; its real-valuedness depends on the adjointness
properties of $\hat{t}$, a question that brings us back to the
physical inner product. While a physical inner product is difficult to
find in such non-deparameterizable situations, one can nevertheless
argue that an imaginary contribution to $\langle\hat{t}\rangle$ in the
system described by (\ref{WdW}) is required in order to provide
equivalence with (\ref{schrod1}).

To do so, we rewrite the right-hand side of (\ref{doubleschrod}) and require
\begin{eqnarray}\label{newop}
\hat{H}^2(\tau,\hat{q},\hat{p})+i\hbar\partial_{\tau}
\hat{H}(\tau,\hat{q},\hat{p})=\hat{H}^2(\hat{t},\hat{q},\hat{p})\,,
\end{eqnarray}
where for distinction we have renamed the parameter $t$ of the Schr\"odinger equation by $\tau$.
Here, $\hat{t}$ is the clock operator in the relativistic system,
which may not be self-adjoint, and $\tau \in\mathbb{R}$ is to be
related to $\langle\hat{t}\rangle$ in some way so as to achieve
equivalence with the Schr\"odinger equation. One can already see
from this equation that imaginary contributions to
$\langle\hat{t}\rangle$ will be required if the left-hand side is
interpreted as some kind of expansion of the right-hand side to
order $\hbar$. In addition to deriving the imaginary contribution,
it remains to be shown that $-\hbar^2\partial_{\tau}^2$ can be
interpreted as $\hat{p}_t^2$, i.e., in terms of the momentum conjugate to the new
operator $\hat{t}$, at least on solutions to (\ref{schrod1}). If
this is the case, (\ref{doubleschrod}) turns into (\ref{WdW}) with
$\hat{t}$ related to $\tau$.

To perform the derivations in the semiclassical approximation, as
sufficient for a comparison with our effective equations, we compute
expectation values of $\hat{H}^2(\hat{t},\hat{q},\hat{p})$ in
solutions to (\ref{schrod1}) assuming the standard Schr\"odinger
inner product up to order $\hbar$.  Then we have $\langle
\hat{t}\rangle^2=\langle \hat{t}^2\rangle$,
$\langle\hat{t}\hat{q}\rangle=
\langle\hat{t}\rangle\langle\hat{q}\rangle$ and
$\langle\hat{t}\hat{p}\rangle=
\langle\hat{t}\rangle\langle\hat{p}\rangle$, just as we have it for
the Zeitgeist associated to $\hat{t}$ of the effective approach with $(\Delta
t)^2=\Delta(t q)=\Delta(t p)=0$; thus,
\begin{eqnarray} \label{Hsquared}
\langle \hat{H}^2(\hat{t},\hat{q},\hat{p})\rangle = \langle
\hat{H}^2(\langle \hat{t} \rangle , \hat{q} , \hat{p}) \rangle +
o(\hbar^{3/2})\,.
\end{eqnarray}
(Even higher-order moments involving $t$ can be expected to vanish,
but equalities here are required only up to order $o(\hbar^{3/2})$.)

We now postulate the relation between $t=\langle\hat{t}\rangle$ and the Schr\"odinger time
$\tau \in{\mathbb R}$ as $t= \tau + i\hbar T$ with $T$ to be
determined. Continuing to expand the right-hand side of
(\ref{Hsquared}), we have
\begin{eqnarray}
\langle \hat{H}^2(t,\hat{q},\hat{p})\rangle &=& \langle
\hat{H}^2(\tau,\hat{q},\hat{p})\rangle + 2i\hbar T
\langle\hat{H}(\tau,\hat{q},\hat{p})\partial_{\tau}\hat{H}(\tau,\hat{q},\hat{p})\rangle
+o(\hbar^{3/2})\nonumber\\
&=&\langle \hat{H}^2(\tau,\hat{q},\hat{p})\rangle+ 2i\hbar
T\langle\hat{H}(\tau,\hat{q},\hat{p})\rangle \langle\partial_{\tau}
\hat{H}(\tau,\hat{q},\hat{p})\rangle+o(\hbar^{3/2})\,.
\label{Hsquared2}
\end{eqnarray}
Combining (\ref{Hsquared}) and (\ref{Hsquared2}), we obtain
(\ref{newop}) in terms of expectation values if
$T=\frac{1}{2\langle\hat{H}\rangle}=\frac{1}{2\langle
i\hbar\partial_{\tau} \rangle}$, the latter equality on solutions of
(\ref{schrod1}).

By construction, recalling (\ref{doubleschrod}), we then have
\begin{eqnarray}\label{almost}
\langle
\hat{H}^2(\hat{t},\hat{q},\hat{p})\rangle=\langle-\hbar^2\partial_{\tau}^2\rangle
\end{eqnarray}
to semiclassical order. For partial time derivatives the imaginary
contribution to $\langle \hat{t} \rangle$ does not matter, and we
may replace $\partial_{\tau}$ by $\partial_t$:
\begin{eqnarray}
\langle
\hat{H}^2(\hat{t},\hat{q},\hat{p})\rangle=\langle-\hbar^2\partial_t^2\rangle=\langle\hat{p}^2_t\rangle\,.
\end{eqnarray}
To semiclassical order solutions to (\ref{schrod1}) satisfy a
relativistic constraint equation if we interpret the expectation
value of the time operator in the latter to be complex with the same
imaginary contribution ${\rm Im}\,\langle \hat{t} \rangle =
-\frac{\hbar}{2\langle\hat{p}_t\rangle}$, as seen in the effective
approach (\ref{imt}).

In terms of operators at a kinematical level, we can identify
\begin{eqnarray}\label{clockop}
\hat{t}=\hat{\tau}-\frac{i\hbar}{2}\widehat{p_{\tau}^{-1}}
\end{eqnarray}
(for states lying outside the zero-eigenspace of $\hat{p}_{\tau}$,
i.e.\ outside ``turning points''). With this identification, we can
further justify replacing $\partial_{\tau}$ by $\partial_{t}$:
thanks to $[\hat{t},\hat{p}_{\tau}]=i\hbar$, the momenta
$\hat{p}_t=\hat{p}_{\tau}$ agree. The Schr\"odinger and relativistic
formulations provide different representations of the dynamics with
different Hilbert spaces. In the representation-independent
effective formulation we have the gauge-fixed constraint
(\ref{Cimag}) as the Schr\"odinger analog, and the non-gauge-fixed
(\ref{Cnon}) as the relativistic analog. In fact, in the toy models
studied in \cite{BHTlongarticle}, the semiclassical dynamics
produced by locally deparametrizing the relativistic constraint with
a Schr\"odinger equation matches precisely the effective dynamics
derived using the corresponding Zeitgeist.

\subsection{Complex time in deparameterizable systems}
\label{s:KGcomplex}

An imaginary contribution to time can be seen also from the well-known
physical inner product formulas available for deparameterizable
systems. An imaginary contribution is not required in those systems
from an effective procedure or for a Schr\"odinger regime, but one can
still see how it may arise naturally.

We consider the free relativistic particle in $1+1$ dimensions,
described by a complex-valued scalar wavefunction of two variables,
$\psi(x_0, x_1)$, subject to the constraint
\begin{equation}
\left( -\hbar^2 \frac{\partial^2}{\partial x_0^2} + \hbar^2
\frac{\partial^2}{\partial x_1^2} - m^2 \right) \psi(x_0, x_1) = 0
\,. \label{eq:free_rel_PDE}
\end{equation}
General solutions have
the form
\begin{equation}
\psi_{\rm phys} (x_0, x_1) = \int_{-\infty}^{\infty} \left( f_+(k)
e^{i\hbar^{-1}(kx_1 - \epsilon_kx_0)} +  f_-(k) e^{i\hbar^{-1}(kx_1
+ \epsilon_kx_0)} \right) {\rm d}x_1 \,, \label{sol_KG_PDE}
\end{equation}
where $\epsilon_k = \sqrt{k^2+m^2}$. Solutions in this general form
automatically split into positive-frequency and negative-frequency
components, a split which is important for constructing the physical
Hilbert space (see, e.g., \cite{GenRepIn}). On positive-frequency
solutions, the physical inner product is
\begin{equation}
\left( \phi, \psi \right) := \left. i\hbar \int_{-\infty}^{\infty}
\left(\bar{\phi}(x_0, x_1) \frac{\partial}{\partial x_0} \psi(x_0,
x_1) - \left( \frac{\partial}{\partial x_0} \bar{\phi}(x_0, x_1)
\right) \psi(x_0, x_1) \right) {\rm d}x_1 \right|_{x_0=t}
\label{eq:KG_prod}
\end{equation}
with an extra minus sign for negative-frequency solutions, while
negative-frequency and positive-frequency solutions are mutually
orthogonal. When evaluated on solutions to (\ref{eq:free_rel_PDE}),
the integration is independent of the value of $t$.

We are interested in an analog of a time operator, which cannot be
an observable. Thus, it does not preserve the space of solutions,
but we can still compute expectation values using (\ref{eq:KG_prod})
as a bilinear form on the kinematical Hilbert space. For
non-observable operators, the expectation values will be time
dependent just as we need it for $t$ itself. For example, for
$\hat{q}=x_1$\ and $\hat{p} = \frac{\hbar}{i}
\frac{\partial}{\partial x_1}$\ the time-dependent expectation
values correspond precisely to the usual dynamics of the free
relativistic particle. Applying this procedure to the time operator, it then becomes apparent that the
expectation value of $\hat{t} = x_0$ (on positive-frequency
solutions $\phi^+$, to be specific) is not only time dependent but
complex:
\begin{eqnarray*}
\left( \phi^+, \hat{t} \phi^+ \right) &=& i\hbar\left.
\int_{-\infty}^{\infty} \left(\bar{\phi}^+(x_0, x_1)
\frac{\partial}{\partial x_0} \left( x_0 \phi^+(x_0, x_1) \right) -
\left( \frac{\partial}{\partial x_0} \bar{\phi}^+(x_0, x_1) \right)
x_0 \phi^+ (x_0, x_1) \right) {\rm d}x_1\right|_{x_0=t} \\&=& i\hbar \left.\int
\bar{\phi}^+ \phi^+ {\rm d}x_1\right|_{x_0=t} + i\hbar \left.\int x_0 \left(
\bar{\phi}^+ \frac{\partial}{\partial x_0} \phi^+ -
\left(\frac{\partial}{\partial x_0} \bar{\phi}^+ \right) \phi^+ \right)  {\rm d}x_1\right|_{x_0=t} \\
&=& i\hbar \left\langle \widehat{\frac{1}{2\epsilon_k}}
\right\rangle + t = t - \frac{i\hbar}{2} \left\langle
\widehat{\frac{1}{p_t}} \right\rangle .
\end{eqnarray*}
(Note that the action of $\hat{p}_t$ on positive-frequency solutions
is equivalent to multiplication by $-\epsilon_k$\ in momentum
space.) Again, to order $\hbar$ the imaginary part of $\langle\hat{t}\rangle$ is in
agreement with the one seen in the effective approach.

For non-deparameterizable systems we do not have an explicit physical inner
product at our disposal, but we can argue heuristically that the time
expectation value should be complex.  We assume a constraint of the
form
\begin{equation}
\left( -\hbar^2 \frac{\partial^2}{\partial x_0^2} - \hat{H}^2(\hat{q},
\hat{p}, \hat{t}) \right) \psi(x_0, x_1) = 0 \,,
\label{eq:general_rel_constr}
\end{equation}
where $ \hat{H}^2$ contains no time derivatives (and thus commutes with
$\hat{t} = x_0$) but may be time-dependent. Solving a second-order
partial differential equation as a constraint, we expect the
physical inner product to depend on both $\psi(x, t)$ and
$\frac{\partial}{\partial t} \psi(x, t)$.  Indeed, it can be shown
that (\ref{eq:KG_prod}) is conserved in time for the solutions of
any constraint of the form given in~(\ref{eq:general_rel_constr}),
so long as $ \hat{H}^2$\ is self-adjoint as an operator on
$L^2(\mathbb{R}, {\rm d}x)$, for each value taken by $t$. However, the
expression is not positive definite in general. It is not difficult
to see, that an inner product involving both $\psi(x, t)$ and
$\frac{\partial}{\partial t} \psi(x, t)$\ will likely assign a
complex expectation value to $\langle \hat{t} \rangle$, since
$\hat{t}$ as a kinematical operator maps $\psi(x_0, x_1)$\ to
$x_0\psi(x_0, x_1)$, and
\begin{equation}
\hat{t}\left( \frac{\partial}{\partial x_0} \psi(x_0, x_1)\right) :=
\frac{\partial}{\partial x_0} (x_0 \psi(x_0, x_1)) = \left(
t\hat{\mathds{1}} + i\hbar \hat{p}_t^{-1} \right)
\frac{\partial}{\partial x_0} \psi(x_0, x_1)\,.
\end{equation}

\subsection{What time is it?}

Confronted with a complex time, we are charged with the task of
elucidating the notion of such a ``vector time'' with its
apparently two separate degrees of freedom. The particular form of
(\ref{imt}) directly implies that the imaginary contribution to the
clock function is a constant of motion in the absence of a
time-dependent potential while it becomes dynamical in the presence
thereof. A constant imaginary contribution can be disregarded
altogether for relational evolution and is also not required in
order to solve the constraints. (\ref{WdW}) and (\ref{schrod1}) are
automatically equivalent in this case; in addition, time does not
appear in the effective constraint functions and may, therefore, be
chosen real. A dynamical imaginary contribution, on the other hand,
can, certainly, not be neglected in the constraints and when
discussing relational evolution. Note that a non-global clock
necessarily implies a time-dependent potential, but a time-dependent
potential does not necessarily imply a non-global clock. For
instance, in a relativistic system with a constraint
$C=p_t^2-H^2(t,q,p)$, where $H^2>0$ $\forall\,t$, the variable $t$
will be a global clock. The dynamical imaginary contribution is,
thus, more general than a mere consequence of non-unitarity, but
becomes most significant where the momentum conjugate to the clock
becomes very small and, accordingly, plays a more pivotal role for
non-global internal times.

This, in fact, also leads us to the discussion of the quality of the
relational clock. For instance, in \cite{GiddingsMarolfHartle} it is
advocated that fundamental uncertainties for relational observables
could arise as a result of using a dynamical variable as clock which
should be disturbed during the measurement of a complete relational
observable. Different clock variables will lead to different
resolutions for relational observables and fundamental uncertainties
could result in general. In \cite{Bianca2}, Poisson brackets of
relational observables are considered from which the uncertainties
will follow in the quantum theory. The inverse kinetic energy of the
clock appears in these Poisson-brackets and it is argued that the
clock is better, the greater its (kinetic) energy,\footnote{Certainly,
large energies are a delicate issue in general relativity, essentially
due to black hole forming, but see \cite{GiddingsMarolfHartle,gampul} and references
therein on this issue in the context of fundamental limits on physical
clocks.} corresponding to the intuition that the faster the clock, the
finer its time-resolution. In agreement with this, it is found in
\cite{aharonovetal} that the quantum notion of a time-of-arrival of a
particle, which in a relational context could be employed as a clock,
is limited by an inherent uncertainty which is inversely proportional
to the kinetic energy of the (clock-)particle. This discussion is in
close parallel to the relations found in this article. The particular
imaginary contribution to the clock is smaller, the larger its kinetic
energy, which is compatible with the fact that the local clock is
better behaved away from its turning point where quantum uncertainties
limit its applicability (see also the following section on this
issue).

As regards relational evolution, we opt to use ${\rm
Re}[\,\langle \hat{t} \rangle]$\ rather than ${\rm Im}[\,\langle \hat{t} \rangle]$\ as the
physical time for several reasons: 1) in the classical limit the imaginary part of $\langle
\hat{t} \rangle$ vanishes and it is, indeed, the real part of $\langle
\hat{t} \rangle$\ that matches the classical time; 2) thanks to
(\ref{imt}), away
from the extremal points, the imaginary part of $\langle \hat{t}
\rangle$\ is small and approximately constant, thus, providing poor
parametrization of dynamics; 3) in the Schr\"odinger regime which linearizes the relativistic constraint, the time parameter refers precisely to the real part
of $\langle \hat{t} \rangle$ (this is, in fact, related to the previous
point as the Schr\"odinger regime is only applicable away from the
extrema); 4) the explicit inner product that reproduces
${\rm Im}[\,\langle \hat{t} \rangle]$\ in the case of a free relativistic
particle is based on integrating at a fixed value of (parameter) $t$\
equal to precisely the real part of the corresponding expectation
value, and 5) the dynamical imaginary term can fail to be monotonic where the real part operates as an appropriate local clock.

\section{Time in a highly quantum state}\label{hq}

\begin{quote}
{\em About the only time we get any let-up from this time control is in the
fog; then time doesn't mean anything. It's lost in the fog, like
everything else.}

\mbox{}\hfill {\sc Ken Kesey}: One flew over the Cuckoo's nest
\end{quote}

Although the specific equations developed here apply only to
semiclassical regimes, general properties of effective constraints
allow us to shed some light on the issue of (non-global) time in
general quantum states. The differences in relational evolution
between the classical and quantum theory merely result, as usual,
from the quantum uncertainties, however, the latter have more severe
repercussions in the absence of a global clock which at the
classical level, in fact, does not constitute a deep conceptual
problem. As always with highly quantum states, intuition becomes
rather foggy; but effective techniques, by being closer in spirit to
the classical formulation than state representations, can provide
valuable input. The role of time in a highly quantum state is a
question of considerable fundamental interest, and it has been
discussed before. Given the difficult nature of this problem,
possible answers put forward so far have remained rather vague.
Proposals derived from the effective constraints in this paper,
expanded semiclassically, will be no less vague. But the viewpoint
they provide is new, we believe, and the light they shed worth
shining.

Recall that at the effective level and to semiclassical order, a
variable can assume the role of a suitable clock wherever its
corresponding Zeitgeist, which fixes all but one effective gauge flow,
is consistent with the assumed hierarchy and fall-off properties of
moments in orders of $\hbar$ as described in section \ref{seceffcon}.
The choice of a Zeitgeist such as (\ref{gauge}), which projects the
clock variable to merely a ``classical'' parameter by setting its
fluctuations to zero,\footnote{It should be noted that due to the
  complex-valuedness of unphysical moments, generalized uncertainty
  relations are respected even when certain fluctuations are zero. For
  instance, in section \ref{secrelho} it was discussed that the
  Zeitgeist (\ref{gauge}) actually leads to a saturation of the
  (generalized) uncertainty relation for the clock variable, a property often associated with a strict form of
semiclassicality.} can be interpreted as
the effective analogue of choosing a constant clock-time slicing in
a deparametrization at the Hilbert-space level which also renders
the clock variable essentially ``classical'', regardless of whether
the state is semiclassical or not.\footnote{The ``classicality" of
an internal time
  variable may be counterintuitive because time, being canonically
  conjugate to a constraint, must be spread out over large domains
  even if it is valid only locally. A state in which time behaves
  semiclassically, by contrast, may be expected to have a sharply
  peaked behavior along the time direction such that time seems unable
  to progress much. The apparent contradiction in the notion of
  semiclassical time is resolved by noting that a wave function
  solving the constraint is indeed spread along the time
  direction, but that semiclassicality must physically be determined
  through properties of the Dirac observables. States are spread out
  kinematically, but time is not an observable and thus lacks obvious
  measures for semiclassicality. (The semiclassicality of other variables, by
contrast, must be derived by solving the constraints and is not
automatically guaranteed.)} In particular, it is really the choice of the clock variable which determines how quantum spreads are measured. Consider, e.g., the deparametrizable example of the free Newtonian particle governed by the constraint $C=p_t+p^2$. Here both $t$ and $q$ are good global clocks and in the quantum theory the physical state solving the quantum version of the constraint will be a priori ``there at once" and infinitely spread in both $t$ and $q$ directions. However, irrespective of how highly quantum the state, we can deparametrize in either $t$ or $q$ by choosing a corresponding slicing on which the physical inner product will be defined. It is this choice of the clock variable which will collapse it to the role of a ``classical" parameter and determine how the spreads of the state are measured, i.e., in this case whether they are measured on a $t=\rm{const}$ or on a $q=\rm{const}$ slice. The clock variable which appears ``classical" in its corresponding slicing might itself appear ``highly quantum" in the slicing corresponding to the other clock choice.

Consider now a system which has no global clock and whose local
clocks have maximal or minimal values along the classical
trajectories. As a prototype of a highly quantum state, consider a
superposition of two or more semiclassical states. For each
classical trajectory, extremal values of a given non-global time
variable are, in general, different, and, therefore, for each of the
corresponding semiclassical states the gauge associated to the clock
choice breaks down at different instants of relational time. For a
superposition of two such states it follows that the region, where a
given time variable is invalid, is larger than for the individual
states. As we superimpose more and more semiclassical states to
obtain a highly quantum solution to the constraint, it is possible,
that, e.g., for systems with closed classical orbits, no regions
remain where a given local time variable can be used as a clock. The
more quantum the state, the more effective variables, i.e.\ higher
moments, and quantum constraint functions we have to take into
account. In such situations it also becomes clear that an analogue
of a gauge associated to the clock, such as the Zeitgeist
(\ref{gauge}) which forces the clock into the role of a ``classical"
parameter, becomes less and less consistent when the quantum nature
of the clock is no longer negligible. In particular, the
fluctuations associated to the momentum conjugate to the clock may
become large as a consequence of superposition of positive and
negative values of the momentum, or, in other words, of opposite
time directions. Note that our construction of the effective
dynamics using local times does not require that the fluctuations of
\emph{all} degrees of freedom can consistently be set to zero or
maintained small, but only of the ones that we want to appoint as
clocks. In addition and related to this, the clock should possess
sufficient kinetic energy, otherwise its resolution is poor and its
imaginary contribution becomes large. If the clock ticks very
slowly, other variables may change significantly in a short interval
of clock time such that their evolution cannot be properly resolved
and fluctuations appear large. Thus, if a highly quantum state has
any degree of freedom that admits a consistent ``projection to a
classical parameter'' and possesses a sufficiently large kinetic energy,
there is a hope that effective dynamics can be defined.

Other methods for defining local time evolution discussed here, fare
no better in a highly quantum state. In general, such a state admits
superpositions of time directions, i.e.\ of positive and negative
frequencies associated to the spectrum of the momentum conjugate to
the clock. This superposition becomes an issue already for
semiclassical states in the turning region of the local clock, where
its conjugate momentum approaches zero, so that both positive and
negative frequencies become relevant. This issue worsens if the spreads are so large that the segments of the wave function before and after the turning region start overlapping. The local Schr\"odinger
regime of section~\ref{schrodreg} relies on using a square-root
operator, which can only be defined on positive or negative
frequency solutions separately. Mixing of the frequencies has the
consequence that we can no longer locally deparametrize in the clock
which would yield a local Schr\"odinger type evolution in only one given time direction
generated by the corresponding Hamiltonian; equivalence of this
regime with the full relativistic constraint, as discussed in
section \ref{schrodreg}, cannot be established anymore and only the
latter is valid. Additionally, in the presence of mixed time
directions, simple inner products based on evaluation at constant
clock-time surfaces seem to be inapplicable and, as a consequence,
it is difficult to see how one could define unitary time evolution
in those cases. As a simple (deparametrizable) example consider once
again the free relativistic particle subject to the constraint
equation~(\ref{eq:free_rel_PDE}) of section~\ref{s:KGcomplex}. This
equation is hyperbolic and the initial value problem (IVP) is a priori
well-posed, but a general solution~(\ref{sol_KG_PDE}) will include
both positive and negative frequencies. Consequently, the
constant-time inner product given by~(\ref{eq:KG_prod}) fails to be
positive-definite and cannot on its own provide us with a physically
meaningful unitary interpretation of the evolution. Only if we
impose the further restriction of only considering, e.g., positive
frequency modes, do we have a positive-definite physical inner
product and a physically meaningful solution to the IVP. The latter
is owed to the fact that restriction to positive frequencies is
tantamount to imposing a (in this case forward pointing) time
direction.\footnote{Also in the classical treatment of relativistic systems, where the square of the momentum conjugate to the clock appears in the constraint,
one is required to specify the time direction in order to formulate a relational IVP. Namely, given the initial data of the other variables at the initial value of the clock, one can only solve the constraint up to sign for the momentum conjugate to the clock. One is forced to choose a sign which then determines the time direction. } It
seems hardly imaginable that, in more general scenarios with
frequency mixings, inner products relying on constant clock-time
surfaces are meaningful. These are usually also closely linked to an
--- at least local --- unitary evolution of initial data in some
clock time, generated by some suitable Hamiltonian. But in a highly
quantum state of a system with no global time even local unitary
evolution becomes meaningless close to the turning region where
frequency mixing is significant --- apart from the fact that
positive and negative frequencies require two separate Hamiltonians
for evolution. A physical inner product based on more general
boundaries or on the entire configuration space is in general
required to cope with such highly quantum scenarios.

Here, however, we rely on local deparametrizations and, therefore, on
disentangling frequencies; at the state level we would like to pose
some IVP at an instant of relational time and at least locally evolve
this initial data unitarily, at the effective level we would like to
impose a gauge such as (\ref{gauge}) and formulate an IVP at a given
clock value only on a segment of the semiclassical orbit, outside the
region where a local clock breaks down, and then evolve data through
this region (using a different local clock). As a result, this
relational concept seems to be of a merely semiclassical nature and
breaks down earlier than the classical evolution in a given clock. The
more quantum the state, the earlier the apparent non-unitarity sets in
and the earlier the relational evolution becomes meaningless. For
sufficiently semiclassical states it is still possible to switch the
clock before non-unitarity sets in,\footnote{In this sense admitting
unitary evolution through the turning point of the clock.} which
amounts to a gauge change in the effective framework and a change of
constant clock-time slicing in local deparametrizations at the state
level. But for highly quantum states this notion of evolution seems to
disappear together with the notion of relational time; if there is no
valid Zeitgeist at the effective level there can also be no
fashionables. This, in fact, is compatible with the breakdown of
relational observables close to turning points in the context of
reduced phase space quantization discussed in \cite{Rovmod}.

The imaginary contribution to time, similarly, is related to local
deparametrizations and constant clock-time slicings; at the
effective level it appears in the gauge associated to the clock
choice and in sections \ref{schrodreg} and \ref{s:KGcomplex} it
showed up in expectation values evaluated in inner products based on
constant time slicings. This complex time might, in fact, obtain
further contributions as we go to higher orders, but, in general,
for an arbitrary quantum state when local deparametrizations and
disentanglement of frequencies are no longer possible, complex time
will disappear together with the notion of relational evolution.

We emphasize that the effects considered here are the result of
imposing a relational interpretation on and attempting a local
reconstruction of physical states of systems without global clocks
and without the usual time structure via local deparametrizations.
The apparent non-unitarity and any decoherence associated to this
are, therefore, a mere result of this interpretation. A priori, the
system may simply lack the standard notion of time-evolution --- and, therefore, of non-unitarity --- altogether.

An arbitrary quantum state will be governed by the full relativistic
constraint and any expectation values of Dirac observables are to be
taken with respect to the physical inner product, which in general
cannot be constructed by evaluating data on a constant clock-time
surface. From the point of view of partial differential equations, it
is hard to see how time-evolution could emerge in general. A general
constraint equation, may not provide a well-posed IVP in any variable
at all. But even if, for a given constraint, the IVP is well-posed on
some constant-time surface, its solution could turn out to be
non-unitary or even non-time-reversible, in the sense that the data at
some later value of relational time is compatible with a multitude of
initial data at the initial surface. Furthermore, such an initial
surface of constant clock time will, in general, intersect the flow
generated by the classical constraint more than once.  Consequently,
assigning initial data on the whole of such a surface lacks clear
physical interpretation as an IVP in the standard sense.  From a
Hilbert-space point of view, it is not clear how to interpret a
general state or distribution (which is after all what one obtains by
solving the constraint) with arbitrary shape/fall off properties as an
at least locally unitary evolution of some sort. Of course, this does
not entirely preclude that there may be a more fundamental way to
define dynamics with respect to some more basic notion of time which
goes beyond the issue of superposition of time directions and reduces
to mere (fuzzy) correlations in a reduced phase space or even Dirac
quantization. However, this remains questionable and even in the
standard relational procedure constructions of quantum relational
observables in the literature remain generally tentative for systems
without global clocks and have otherwise only been successfully
completed in the deparametrizable case. In contrast to this, the
advantage of the effective approach is that it naturally gives rise to
the notion of fashionables semiclassically and offers an outlook to
more quantum regimes, suggesting that the nature of time changes as
one motions from classical behavior to more highly quantum states.

\section{Discussion}

We have applied the effective procedure of dealing with quantum
constraints to non-deparameterizable systems. Traditional procedures
to deal with physical evolution are difficult to apply for those
systems, but the effective approach is very feasible, at least for
semiclassical questions which are often of most interest in those
aspects of quantum gravity or cosmology that have at least a slight
chance of being potentially observable.  Within the same approach, not
only solving the constraints but also finding relational observables
benefits from strong simplifications compared to calculations for
operators and states in a Hilbert-space representation. Computing
explicit observables may still be complicated, but no conceptual
problems occur and numerical tools can easily be implemented.

In particular, many of the facets of the problem of time are evaded by
the possibility of patching together local internal times in quantum
systems, just as one could do it for classical systems. Physicality
conditions for observables are implemented just by reality conditions;
no integral representation of an inner product need be
constructed. Quantizing non-deparameterizable systems often fails
already at the step of constructing a physical inner product, a strong
handicap for canonical quantum gravity which is completely avoided by
the effective techniques. Even if one were to know a physical inner
product, finding sufficiently many quantum observables is often a
problem.  This step as well is simplified in the effective formalism
which is treated in a classical manner and is also more amenable to
numerical implementations. With the methods described here, we, thus,
expect that much headway can be made in evaluating quantum-gravity
theories and models in a practical way. Obviously, the effective
setting remains to be extended, most importantly to quantum field
theories, before it becomes applicable to full quantum gravity in
semiclassical regimes. Promisingly, in formulations such as loop
quantum gravity one can replace the continuous field theories of
gravity by systems of finitely many degrees of freedom in compact
regions, for instance by focussing attention on suitable classes of
spin-network states which still capture the full amount of degrees of
freedom. Therefore, crucial issues of quantum field theory should not
be expected to make the effective techniques for constrained systems
inapplicable in the context of gravity in inhibiting ways.

The main advantage for non-deparameterizable systems is that local
internal times can be made sense of in the first place, and patched
together by simple gauge changes. With changes of complex local
internal times shown to be fully consistent, patching local internal
times becomes a valid procedure, overcoming the (global and multiple
choice) problem of time. For any concrete system, one will notice
the breakdown of one's initial choice of internal time when its
momentum becomes small, approaching a turning point of time. Before
the momentum takes on too small values, which would endanger the
validity of the semiclassical approximation, one can transform to a
new internal time. No sharp instant can be provided for when the
change of time should be performed, but it is not relevant as long
as the change of time is performed well before the breakdown of the
initial choice of time \cite{BHTlongarticle}.

The most striking feature of non-global clocks is their necessary
complex-valuedness which can be elucidated by general arguments.
Regarding evolution, we advocate in this article to appoint only the
real part as relational time. Local internal times, furthermore,
allow one to extend the notion of quantum relational observables to
non-deparameterizable systems. Fashionables such as
$\langle\hat{q}\rangle(\langle\hat{t}\rangle)$\ are state-dependent and result even if
$\langle\hat{t}\rangle$\ is not used as internal time throughout the
whole evolution. However, local internal times imply several
subtleties, and for this reason the notion of fashionables is more
general than the one of relational observables developed mainly with
deparameterizable systems in mind. For instance, we are obliged to
perform a change of time and associated gauge prior to the turning
point of the clock, resulting in (order $\hbar$) discontinuities in
correlations and the necessity to evolve a truly {\it different} set
of fashionables in the new Zeitgeist. The relational concept is of a
genuinely local nature here.

By sidestepping the Hilbert space problem, the effective approach
also offers a vague outlook on the nature of time in highly quantum
states, indicating that the usual concept of evolution in a
relational time disappears in highly quantum regimes of systems
devoid of global clocks.

For actual physical predictions of a theoretical framework,
relational Dirac observables are not required to be defined or known
for all values of relational time. Observations refer only to finite
ranges of time, and so for predictions relational observables need
be known only for finite ranges of relational time. These finite
ranges may not even be contiguous. For instance, one may be
interested in the change of an observable as the variable used as
internal time moves through a turning point. (A cosmological
application may be the evolution through a bounce.) Differences in
Zeitgeist do not pose problems for these questions; one could simply
disregard any gaps in the complete relational evolution as well as
in differences of gauge for intermediate periods bringing one
through the turning point of an internal time of interest. For
evolution before and after the turning point the same local internal
time can be used, removing potential interpretational difficulties
that could arise from the use of different choices of Zeitgeist.

For further discussion of the issues raised in this article and
concrete examples, we refer the interested reader to
\cite{BHTlongarticle}.

\section*{Acknowledgements}

It is a pleasure to thank Igor Khavkine and Renate Loll for
interesting discussions and Bianca Dittrich for useful comments on an earlier version of this manuscript. This work was supported in
part by NSF grant PHY0748336 and a grant from the Foundational
Questions Institute (FQXi). PAH is grateful for the support of the
German Academic Exchange Service (DAAD) through a doctoral research
grant and acknowledges a travel grant of Universiteit Utrecht.
Furthermore, he would like to express his gratitude to the Albert
Einstein Institute in Potsdam for hospitality during the final
stages of this work.


\end{document}